\documentclass[letter,twocolumn]{jpsj3} 
\usepackage{ulem}
\usepackage{color}

\title{
Doping Effects on the Electronic Structure of an Anisotropic Kondo Semiconductor CeOs$_2$Al$_{10}$:
An Optical Study with Re and Ir Substitution
}
\author{
Shin-ichi Kimura$^{1,2}$\thanks{E-mail address: kimura@fbs.osaka-u.ac.jp}, 
Hitoshi Takao$^2$,
Jo Kawabata$^3$,
Yoshihiro Yamada$^3$,\\
and 
Toshiro Takabatake$^{3,4}$
}
\inst{
$^1$Graduate School of Frontier Biosciences, Osaka University, Suita, Osaka 565-0871, Japan\\
$^2$Department of Physics, Osaka University, Toyonaka, Osaka 560-0043, Japan\\
$^3$Department of Quantum Matter, ADSM, Hiroshima University, Higashi-Hiroshima, Hiroshima 739-8530, Japan\\
$^4$Institute for Advanced Materials Research, Hiroshima University, Higashi-Hiroshima, Hiroshima 739-8530, Japan\\
}
\recdate{
October 19, 2016, Accepted November 2, 2016
}
\abst{
An anisotropic Kondo semiconductor CeOs$_2$Al$_{10}$ exhibits an unusual antiferromagnetic order at rather high transition temperature $T_0$ of 28.5~K.
Two possible origins of the magnetic order have been proposed so far, one is the Kondo coupling of the hybridization between the conduction ($c$) and the $4f$ states and the other is the charge-density wave/charge ordering along the orthorhombic $b$ axis.
To clarify the origin of the magnetic order, we have investigated the electronic structure of hole- and electron-doped CeOs$_2$Al$_{10}$ [Ce(Os$_{1-y}$Re$_y$)$_2$Al$_{10}$ and Ce(Os$_{1-x}$Ir$_x$)$_2$Al$_{10}$, respectively] by using optical conductivity spectra along the $b$ axis.
The intensity of the $c\mathchar`-f$ hybridization gap at $\hbar\omega\sim50$~meV continuously decreases from $y=0.10$ to $x=0.12$ via $x=y=0$.
The intensity of the charge excitation observed at $\hbar\omega\sim20$~meV has the maximum at $x=y=0$ as similar with the doping dependence of $T_{\rm 0}$.
The fact that the charge excitation is strongly related to the magnetic order strengthens the possibility of the charge density wave/charge ordering as the origin of the magnetic order.
}
\begin{document}
\maketitle
%
%
Recently, materials lacking local inversion symmetry with strong spin-orbit coupling are attracting attention 
because of their various physical properties, 
such as superconductivity with a very high critical magnetic field~\cite{Bauer2004} and
extreme-high spin polarization of surface/bulk Rashba states.~\cite{Rashba}
The title compound, CeOs$_2$Al$_{10}$, can be regarded as one of materials with lacking local inversion symmetry.

The compounds Ce$M_2$Al$_{10}$ ($M=$~Os and Ru) are Kondo semiconductors (KS's) with a small energy gap at low temperature.~\cite{Strydom2009}
The crystal structure is orthorhombic YbFe$_2$Al$_{10}$-type,~\cite{Thiede1998} in which the Ce-$M$ bonding forms a zigzag plane in the $ac$ plane,~\cite{Kimura2011-2}
although so called Kondo insulators (KIs) such as SmB$_6$ and YbB$_{12}$ have cubic crystal structure.~\cite{Takabatake1998}
In the compounds with $M=$~Os and Ru, an antiferromagnetic order appears at the temperature of about 28~K that is lower than the Kondo temperature of about 60~K.~\cite{Nishioka2009}
The ordering temperature is much higher than that expected by the de Gennes factor.~\cite{Muro2011}
The origin of the high ordering temperature has not been clarified yet.
One possible origin is considered to be the low crystal symmetry and/or the lacking of local inversion symmetry.

To clarify the novel physical properties of Ce$M_2$Al$_{10}$s, we have investigated the electronic structure by measuring polarized optical conductivity [$\sigma(\omega)$] spectra.~\cite{Kimura2011-2,Kimura2011-1,Kimura2011-3}
So far, we have revealed the anisotropic electronic structure and the anisotropic hybridization intensity between conduction ($c$) and $4f$ states ($c\mathchar`-f$ hybridization), 
i.e., strong (weak) $c\mathchar`-f$ hybridization in the $ac$ plane (along the $b$ axis).
Especially, we could find that the $c\mathchar`-f$ hybridization gap ($\Delta_{c\mathchar`-f}$) with the size of about 50~meV appears along all axes.
In addition to this, we have found that a charge gap ($\Delta_0$) with the size of 20~meV appears only along the $b$ axis.~\cite{Kimura2011-1,Kimura2011-2}
Since the $\Delta_F0$ does not appear in non-magnetic CeFe$_2$Al$_{10}$ and is formed above $T_0$ in CeOs$_2$Al$_{10}$ (38~K) and CeRu$_2$Al$_{10}$ (30~K), the presence of $\Delta_0$ seems to be related to the magnetic order,
i.e., the electronic structure along the $b$ axis plays an important role for the magnetic order.

The relation between the $c\mathchar`-f$ hybridization gap and the magnetic order in CeRu$_2$Al$_{10}$ has been investigated by using the Rh-substituted materials [Ce(Ru$_{1-x}$Rh$_x$)$_2$Al$_{10}$].~\cite{Kimura2015}
The paper claimed that the intensity of $\Delta_{c\mathchar`-f}$ is strongly suppressed in between $x=0.03$ and $0.05$, suggesting that the $c\mathchar`-f$ hybridization intensity drastically decreases.
This result is consistent with the other results that the $4f$ electron is well localized and the $c\mathchar`-f$ hybridization strength becomes weak by the 5~\% substitution of Rh for Ru ($x=0.05$).~\cite{Kondo2013,Tanida2014-2,Tanida2016}
On the other hand, with increasing $x$, $T_0$ monotonically decreases from 27.3~K to 24~K for $x=$~0.05 and disappears at $x \sim 0.3$~\cite{Kobayashi2013}.
The intensity of $\Delta_0$ hardly decreases and still remains at $x=0.05$.~\cite{Kimura2015}
This change in the intensity of $\Delta_0$ is in parallel with that of $T_0$.
However, since not only $T_0$ but also both of the $c\mathchar`-f$ hybridization intensity and the carrier density monotonically change with $x$, the origin of the magnetic order cannot be concluded to be the appearance of $\Delta_0$.
In addition, the recent tunneling spectroscopic study of $5d$-electron doped CeOs$_2$Al$_{10}$ claimed that the appearance of the magnetic order needs the $c\mathchar`-f$ hybridization gap.~\cite{Kawabata2015,Kawabata2017}

In this Letter, we report the relation among $T_0$, $\Delta_0$, $\Delta_{c\mathchar`-f}$, and the carrier density in the wide substitution range from the hole-doped [Ce(Os$_{1-y}$Re$_y$)$_2$Al$_{10}$] to the electron-doped [Ce(Os$_{1-x}$Ir$_x$)$_2$Al$_{10}$] materials~\cite{Kawabata2014} by using the $\sigma(\omega)$ spectra.
As a result, the $c\mathchar`-f$ hybridization intensity gradually decreases with decreasing $y$ and increasing $x$, while the intensity of $\Delta_0$ also takes maximum at $x=y=0$ and decreases with increasing $x$ and $y$.
This dependence of $\Delta_0$ is roughly in parallel to that of $T_0$.
On the other hand, the carrier density behaves in the opposite to $\Delta_0$.
Therefore, the magnetic order is strongly related to $\Delta_0$ and the carrier density.

%
\begin{figure}[t]
\begin{center}
\includegraphics[width=0.45\textwidth]{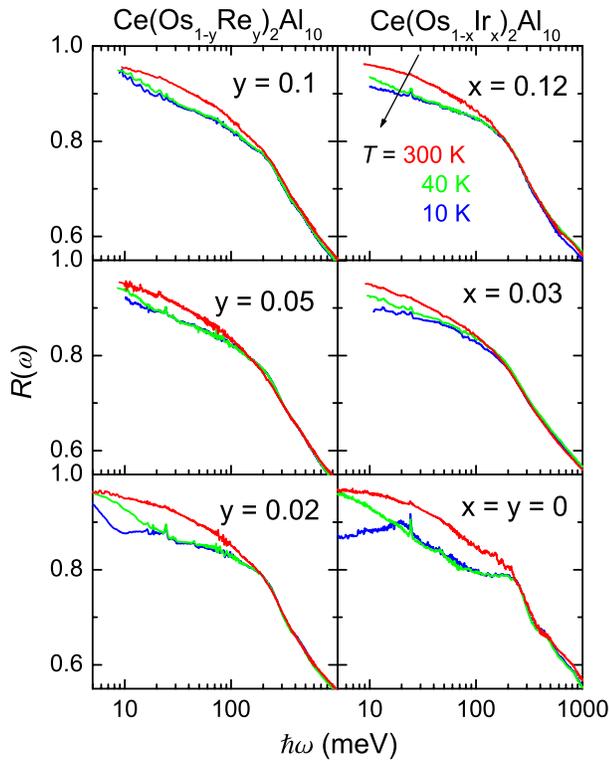}
\end{center}
\caption{
(Color online)
Temperature-dependent reflectivity [$R(\omega)$] spectra of Ce(Os$_{1-y}$Re$_y$)$_2$Al$_{10}$ ($y=0.02, 0.05, 0.1$) and Ce(Os$_{1-x}$Ir$_x$)$_2$Al$_{10}$ ($x=0.03, 0.12$) along the $b$ axis.
The $R(\omega)$ spectra of $x=y=0$ are referred from Ref.~\citen{Kimura2011-1}.
}
\label{fig:reflectivity}
\end{figure}
Single-crystalline samples of [Ce(Os$_{1-y}$Re$_y$)$_2$Al$_{10}$, $y=0.02, 0.05, 0.10$, Ce(Os$_{1-x}$Ir$_x$)$_2$Al$_{10}$, $x=0.03, 0.12$] were synthesized by the Al-flux method.~\cite{Muro2010-2}
The composition was determined by the wavelength dispersion electron-probe microanalysis.~\cite{Kawabata2015}
The surfaces were well-polished using 0.3~$\mu$m grain-size Al$_{2}$O$_{3}$ lapping film sheets for the optical reflectivity [$R(\omega)$] measurements.
Near-normal incident polarized $R(\omega)$ spectra were acquired in a very wide photon-energy range of 2~meV -- 30~eV to ensure an accurate Kramers-Kronig analysis (KKA)~\cite{Kimura2013}.
Martin-Puplett and Michelson type rapid-scan Fourier spectrometers (JASCO Co. Ltd., FARIS-1 and FTIR6100) were used at the photon energy $\hbar\omega$ regions of 2~--~30~meV and 5~meV~--~1.5~eV, respectively, with a feed-back positioning system to maintain the overall uncertainty level less than $\pm$0.5~\% in the temperature range of 10~--~300~K~\cite{Kimura2008}.
Samples were mounted in a high-vacuum cryostat ($\leq10^6$~Pa) that includes in-situ evaporation system to cover sample with a reference gold film.

The obtained polarized $R(\omega)$ spectra of all samples in the photon energy range of 5~meV -- 1~eV are plotted in Fig.~\ref{fig:reflectivity} with those of CeOs$_2$Al$_{10}$ ($x=y=0$) referred from Ref.~\citen{Kimura2011-1}.
At $T=300$~K, $R(\omega)$ of CeOs$_2$Al$_{10}$ was only measured for energies 1.2--30~eV by using synchrotron radiation (not shown here), and connected to all spectra for the KKA.
In order to obtain $\sigma(\omega)$ via KKA of $R(\omega)$, the spectra were extrapolated below 2~meV with a Hagen-Rubens function, and above 30~eV with a free-electron approximation $R(\omega) \propto \omega^{-4}$~\cite{DG}.


\begin{figure}[t]
\begin{center}
\includegraphics[width=0.45\textwidth]{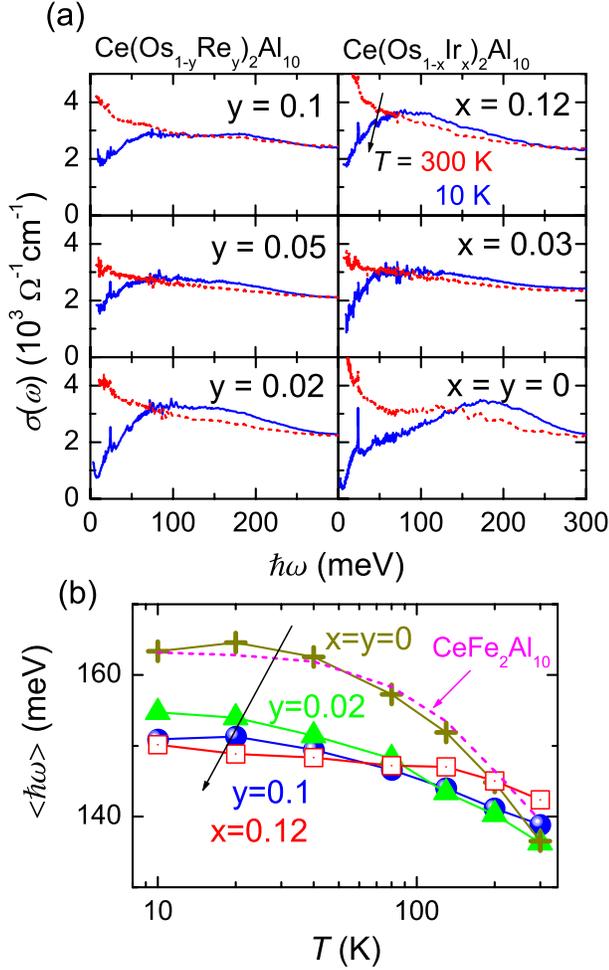}
\end{center}
\caption{
(Color online)
(a) Temperature-dependent optical conductivity [$\sigma(\omega)$] spectra of Ce(Os$_{1-y}$Re$_y$)$_2$Al$_{10}$ ($y=0.1, 0.05, 0.02$), Ce(Os$_{1-x}$Ir$_x$)$_2$Al$_{10}$ ($x=0.12, 0.03$), and $x=y=0$ along the $b$ axis in the photon energy $\hbar\omega$ region below 300~meV.
(b) The centers of gravity [$\langle\hbar\omega\rangle$] of the $\sigma(\omega)$ spectra in $\hbar\omega\leq$~300~meV of $x=$~0.02 (solid triangle), 0.1 (solid circle), $x=0.12$ (open square), $x=y=0$ (cross), and CeFe$_2$Al$_{10}$ (dashed line)~\cite{Kimura2011-3} are plotted as a function of temperature $T$.
The error of $\langle\hbar\omega\rangle$ is less than the size of marks.
}
\label{fig:OCwide}
\end{figure}
The $\sigma(\omega)$ spectra for $\hbar\omega\leq$~300~meV are shown in Fig.~\ref{fig:OCwide}(a).
The wide-range $R(\omega)$ spectrum in Fig.~\ref{fig:reflectivity} as well as the $\sigma(\omega)$ spectrum hardly changes on cooling in the photon energy range of $\hbar\omega>$~300~meV.
Therefore the $\sigma(\omega)$ spectrum has temperature dependence only in the range of $\hbar\omega\leq$~300~meV.
A broad peak observed at about 170~meV (mid-IR peak) in $x=y=0$ originates from the optical transition from the valence band to the unoccupied Ce~$4f_{5/2}$ state~\cite{Kimura2009}.
Both the peak intensity and the temperature dependence, which denote the $c\mathchar`-f$ hybridization character, 
imply that the $c\mathchar`-f$ hybridization intensity decreases with increasing $x$ and $y$~\cite{Kimura2016}.
To clarify the temperature dependence, the center of gravity ($\langle\hbar\omega\rangle$)
below 300~meV 
is plotted in Fig.~\ref{fig:OCwide}(b).
The temperature dependence of $\langle\hbar\omega\rangle$ for $x=y=0$ is the largest even though that at room temperature is comparable with those of doped samples.
The temperature dependence is strongly suppressed by doping of holes and electrons, indicating the rapid collapse of the $c\mathchar`-f$ hybridization band structure.
Therefore we point out that this system is described by the Kondo lattice model rather than the single-site Kondo coupling model.

To compare the $\langle\hbar\omega\rangle$ of CeOs$_2$Al$_{10}$ with that of CeFe$_2$Al$_{10}$, 
$\langle\hbar\omega\rangle$ of CeFe$_2$Al$_{10}$ has been calculated using the data in Ref.~\citen{Kimura2011-3} and plotted in Fig.~\ref{fig:OCwide}(b).
The temperature dependence of $\langle\hbar\omega\rangle$ for $x=y=0$ is similar to that of CeFe$_2$Al$_{10}$ but slightly shift to the lower temperature side.
Since the $\langle\hbar\omega\rangle$ indicates the temperature dependence of the Kondo coupling intensity,~\cite{Iizuka2010}
this is qualitatively consistent with that the Kondo temperature of CeOs$_2$Al$_{10}$ ($\sim60$~K) is lower than that of CeFe$_2$Al$_{10}$ ($\sim80$~K).~\cite{Nishioka2009,Kimura2012}

\begin{figure}[t]
\begin{center}
\includegraphics[width=0.45\textwidth]{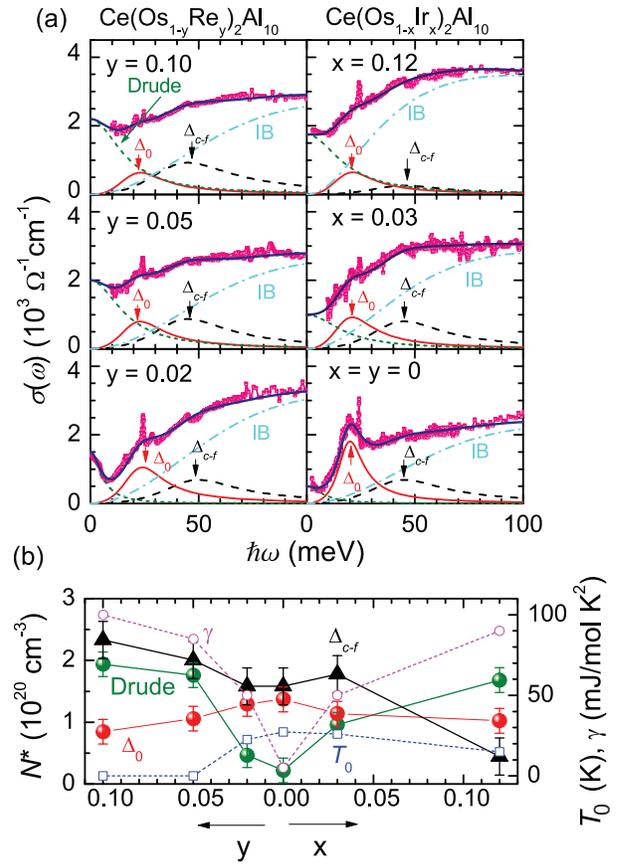}
\end{center}
\caption{
(Color online)
(a) $\sigma(\omega)$ spectra along the $b$ axis at $T=10$~K and the fitted lines of one Drude and three Lorentz functions.
The three Lorentz functions are assumed as a peak ($\Delta_0$, solid line) appearing below the magnetic ordering temperature $T_0$, the interband transition in the $c\mathchar`-f$ hybridization gap ($\Delta_{c\mathchar`-f}$, dashed line), and higher interband transition from lower energy side (IB, dot-dashed line).
(b) Obtained effective electron numbers ($N^*$s) of the Drude, $\Delta_0$, and $\Delta_{c\mathchar`-f}$ components, which are plotted together with $T_0$ and the electronic specific heat coefficient $\gamma$ as functions of the Re ($y$) and Ir ($x$) concentrations.~\cite{Kawabata2014}
}
\label{fig:OCnarrow}
\end{figure}
Next, we focus on the energy gap structure in the electronic structure near the Fermi level ($E_{\rm F}$).
The $\sigma(\omega)$ spectra along the $b$ axis at $T=$~10~K in $\hbar\omega\leq$~100~meV are shown in Fig.~\ref{fig:OCnarrow}(a).
In this region, it should be noted that a very sharp peak owing to a transverse optical phonon is observed at around 25~meV in all samples.
The temperature and concentration dependences of the peak will be discussed in a separated paper because the peak is not directly related to the electronic structure.

To clarify the change of the spectra by increasing $x$ and $y$, we adopt the Drude-Lorentz fitting method that has been useful in analyzing the data for Ce(Ru$_{1-x}$Rh$_x$)$_2$Al$_{10}$.~\cite{Kimura2015}
The $\sigma(\omega)$ spectra are fitted by using the combination of Lorentz functions:
\[
\sigma(\omega)=\sum_i\frac{N^*_ie^2\tau_i\omega^2}{m_e[(\omega^2_i-\omega^2)^2\tau^2_i+\omega^2]}.
\]
Here, $i$ has four components, Drude ($i=D$) with $\omega_D=0$, $\Delta_0$ (0), $\Delta_{c\mathchar`-f}$ ($c\mathchar`-f$), and a higher interband transition (IB).
$m_e$ denotes the electron rest mass.
$N^*_i$, $\tau_i$, and $\omega_i$ are the effective electron number, the relaxation times, and the resonant frequency of the Lorentz absorption of each component, respectively.
The fitting was performed by using a least square method with keeping the peak center energies of $\hbar\omega_0$ and $\hbar\omega_{c\mathchar`-f}$ at 20--23~meV and 45--48~meV, respectively.
Here, we assumed that the gap energies and shapes are reasonably constant against doping.
The IB curves were fitted to reproduce the slope at $\hbar\omega=$~100--120~meV.
The obtained values of the effective electron number $N^*_i$ together with $T_0$ and $\gamma$ are plotted in Fig.~\ref{fig:OCnarrow}(b) as functions of Re ($y$) and Ir ($x$) concentrations.~\cite{Kawabata2014}

In Figure~\ref{fig:OCnarrow}(a), the $\Delta_0$ peak of $x=y=0$ is narrower than those of others.
The reason is not clear but the corresponding peak in a tunneling spectrum is also narrower than those of doped materials.~\cite{Kawabata2017}
Therefore, the peak shape might be reasonable and the origin should be clarified in future.

From Figure~\ref{fig:OCnarrow}(b), we can get useful informations as follows:

Firstly, the doping dependence of $N^*$s of the Drude component ($N^*_D$) is qualitatively consistent with that of $\gamma$ value.
This is reasonable because both of the $N^*_D$ and the $\gamma$ value are proportional to the density of states at $E_{\rm F}$.
However we can recognize that the $N^*_D$ at $y=0.02$ is much smaller than that expected from the $\gamma$ value in comparison with other concentrations.
The deviation of the $N^*_D$ from the $\gamma$ value at $y=0.02$ might originate from the mass enhancement owing to the strong electron correlation at low carrier density
because $N^*_D=N m_e/m^*$, where $N$ and $m^*$ are the carrier density and the effective mass of carriers, respectively, 
and $N$ must be proportional to the $\gamma$ value.

Secondly, the $N^*$ of $\Delta_{c\mathchar`-f}$ gradually decreases with decreasing $y$ and increasing $x$ within error bars.
This is consistent with the doping dependence of magnetic susceptibility,~\cite{Kawabata2014} i.e., 
the $4f$ state is more localized (the $c\mathchar`-f$ hybridization strength decreases) with increasing electron carriers.
This result is also consistent with the results obtained for the Rh doping in CeRu$_2$Al$_{10}$.~\cite{Kimura2015}
However, the doping-dependent behavior of the $c\mathchar`-f$ hybridization gap is inconsistent with the behavior of the mid-IR peak shown in Fig.~\ref{fig:OCwide}.
The reason is not clear at present, but two reasons as follows are proposed: 
One is that the effective probing area of the $c\mathchar`-f$ hybridization at the mid-IR peak might be different from that near $E_{\rm F}$, i.e.,
the periodicity of the $c\mathchar`-f$ hybridization can be probed by the mid-IR peaks in contrast that 
the average of the $c\mathchar`-f$ hybridization intensity appears in $\Delta_{c\mathchar`-f}$ because the photon energy as well as the wavelength is much different.
The other is the screening effect by carriers, i.e., the mid-IR peak is more sensitive to the carrier density than $\Delta_{c\mathchar`-f}$.

Lastly, the doping dependence of the $N^*$ of $\Delta_0$ ($N^*_0$) is qualitatively consistent with that of $T_0$.
Therefore we can confirm the previous argument that the electronic structure producing the $\Delta_0$ peak is strongly related to the magnetic order at $T_0$.
In addition, since the $N^*_0$ at $y=0.05$ and $0.1$ is not zero, the presence of $\Delta_0$ is not sufficient for the magnetic order.
In CeOs$_2$Al$_{10}$ as well as CeRu$_2$Al$_{10}$, $\Delta_0$ starts to appear at slightly higher temperature ($T^*$) than $T_0$.~\cite{Kimura2011-1,Kimura2011-2}
At temperatures between $T^*$ and $T_0$, no magnetic ordering appears.
The same situation might appear in $y=0.05$ and $0.1$.
From the figure, we can recognize that the $N^*_D$ has an opposite behavior to the $N^*_0$.
This suggests that the appearance of the $\Delta_0$ is related to the carrier density, i.e., 
the electronic structure of $\Delta_0$ appears when the carrier density decreases by the gap opening owing to the strong $c\mathchar`-f$ hybridization of KS.
This behavior is similar to that of charge-density wave (CDW) materials~\cite{DiSalvo1975} as well as the charge ordering (CO) material Yb$_4$As$_3$.~\cite{Kimura1996}

In KIs as well as KS's, the appearance of a CDW as well as a CO phase was theoretically expected by using the periodic Anderson model with the electron correlation between carriers and $4f$ electrons, $U_{c\mathchar`-f}$, of proposed by Yoshida and coworkers.~\cite{Yoshida2011,Yoshida2012}
In Ce$M_2$Al$_{10}$ ($M=$ Os, Ru), the $\Delta_0$ starts to open at slightly higher temperature than $T_0$~\cite{Kimura2011-1, Kimura2011-3}, which can be explained by their calculation.~\cite{Yoshida2011}
According to the theory, the CDW/CO phase appears in the high $U_{c\mathchar`-f}$ region apart from the antiferromagnetic phase.
The RKKY interaction as well as a magnetic order temperature is enhanced by increasing the $c\mathchar`-f$ hybridization intensity owing to the renormalization effect of the $U_{c\mathchar`-f}$.~\cite{Yoshida2012}
Therefore the magnetic order might occur at rather high temperatures in the presence of the CDW/CO phase.

However, in general, the effect of $U_{c\mathchar`-f}$ as well as the CDW/CO phase does not appear in three-dimensional KIs such as SmB$_6$ and YbB$_{12}$, 
but only appear in a low carrier system, Yb$_4$As$_3$.~\cite{Ochiai}
The crystal structure of the latter can be regarded as composed of one-dimensional Yb chains along the $\langle111\rangle$ axis.~\cite{Antonov1998}
The CO peak of Yb$_4$As$_3$, which appears below slightly higher temperature than CO,~\cite{Kimura2002} shows the similar temperature dependence to that of the $N^*_0$ of Ce$M_2$Al$_{10}$.
Therefore the effect of $U_{c\mathchar`-f}$ is regarded to appear also in Ce$M_2$Al$_{10}$.
Because of no magnetic order in Yb$_4$As$_3$, however, the high-temperature magnetic order in Ce$M_2$Al$_{10}$ is considered to be a novel physical phenomenon owing to the $U_{c\mathchar`-f}$ and the anisotropic crystal structure.

%
In conclusion, the $5d$ hole and electron doping effects on the electronic structure of Ce(Os$_{1-y}$Re$_y$)$_2$Al$_{10}$ ($y=0.02, 0.05, 0.1$) and Ce(Os$_{1-x}$Ir$_x$)$_2$Al$_{10}$ ($x=0.03, 0.12$) have been investigated by using optical conductivity spectroscopy.
The absorption intensity of the charge gap at $\hbar\omega\sim20$~meV changes roughly proportional to $T_0$ in spite that the absorption intensity of the $c\mathchar`-f$ hybridization gap at $\hbar\omega\sim50$~meV monotonically decreases with increasing the electron dope level.
The evaluated effective electron number of the Drude component is inversely proportional to the absorption intensity of the charge gap.
These results suggest that the charge gap as well as the magnetic order appear at the low carrier density owing to the weak screening effect of carriers.
For better understanding the obtained phenomena, theoretical studies with the periodic Anderson model with the correlation between carriers and $4f$ electrons are highly anticipated.

\section*{Acknowledgments}
We would like to acknowledge H. Tanida, M. Sera, Y. Muro, A. Kondo, and T. Yoshida for their fruitful discussion and H. Yokoyama and S. Kamei for their support during optical experiments.
We also thank UVSOR staff members for their technical support.
Part of this work was performed under the Use-of-UVSOR Facility Program (BL1B, 2014-2016) of the Institute for Molecular Science.
This work was partly supported by JSPS KAKENHI (Grant Nos. 26400363 and 16H01076).


%
\end{document}